# Theoretical Remarks on Cybotactic Clusters of Bent-Core Nematic Liquid Crystals in 1D Settings

Sourav Patranabish [a], Yiwei Wang [b], Aloka Sinha [a] and Apala Majumdar [c]

[a] Department of Physics, Indian Institute of Technology Delhi, Hauz Khas, New Delhi – 110016, India
[b] School of Mathematical Sciences, Peking University, Beijing 100871, China
[c] Department of Mathematical Sciences, University of Bath, Bath BA2 7AY, UK

**Abstract**

The bent-core liquid crystals (LCs) are highly regarded as the next-generation materials for electro-optic devices. The nematic (N) phase of these LCs possesses highly ordered smectic-like cybotactic clusters which are promising in terms of ferroelectric-like behaviour in the N phase itself. We have studied a one-dimensional (1D) Landau-de Gennes model of spatially inhomogeneous order parameters for the N phase of bent-core LCs. We investigate the effects of spatial confinement and coupling (between these clusters and the surrounding LC molecules) on the order parameters to model cluster formation in recently reported experiments. The coupling is found to increase the cluster order parameter significantly, suggesting an enhancement in the cluster formation and could also predict a possible transition to a phase with weak nematic-like ordering in the vicinity of nematic-isotropic transition upon appreciable increase of the coupling parameter $\gamma$.

## 1. Introduction

Bent-core liquid crystals (LCs) are regarded as promising materials for the next-generation of electro-optic devices and applications [1,2]. A LC molecule usually has a rod-like shape with a uniaxial cylindrical symmetry [3]. When a bent-unit is introduced between two such rod-like arms of an LC molecule, a bent-core LC molecule can be realized, but now with a reduced symmetry of $C_{2v}$ type [3]. The bent-core LCs are a recent inclusion in the family of liquid crystals, offering wide spread applications in LC based devices [4]. Ferroelectric LCs are identified in terms of the spontaneous polarization ($P_S$) existing in an LC system that follows the externally applied time varying AC electric-field in a periodic fashion [5]. The fast and flicker-free high resolution ferroelectric liquid crystal displays (FLCDs) have several advantages over the conventional liquid crystal displays (LCDs) including very fast microsecond (μs) switching and bistability for better video/picture frame rates [5,6]. The first ferroelectric LC was discovered in the chiral smectic C (SmC*) phase (with a reduced symmetry from $C_{2h}$ to $C_2$) [3], but the bent-core LCs offer the possibility of ferroelectric-like behaviour in the nematic (N) phase itself with even faster switching for next-generation displays [2,7–10]. The nematic (N) phase of bent-core LCs, equipped with the possibility to exhibit macroscopic biaxiality [11–13] and the elusive ferroelectric nematic ($N_f$) phase [2], has promising applications in electro-optic devices with added advantages [3,4,9].

The N phase of calamitic LC compounds commonly possesses long-range orientational order, but lack in any definite positional correlation or ordering. In contrast, the N phase exhibited by majority of the bent-core LCs, apart from the usual long-range orientational ordering, possess an additional short-range smectic-like local positional ordering, termed as the cybotactic clusters [2,4,7–10]. It is believed that the bent-shaped LC molecules owing to their kinked molecular shape, lock themselves together to form a cybotactic cluster with short-range orientation and positional correlation [14]. The highly debated cybotactic nematic phase ($N_{Cyb}$) of bent-core mesogens has been studied extensively over the past few years and is now established as a reality with concrete evidence [2,15]. Although the $N_{Cyb}$ phase was initially considered to be a pre-transitional effect arising only in the vicinity of the N to smectic (Sm) phase transition, experiments have now revealed their existence in the whole N phase, even in absence of an underlying Sm phase of the compounds [15,16]. Recent experimental studies show that these short-range Sm-like cybotactic clusters can exist even in the isotropic liquid phase of the bent-shaped compounds [9,14]. Owing to the highly ordered microstructure of these clusters, the $N_{Cyb}$ phase of bent-shaped LCs becomes an excellent candidate for exhibiting macroscopic biaxiality and the ferroelectric nematic ($N_f$) phase [7,8,10,11,17,18]. Studies have also shown that a strong coupling between the polar order and the director gradients can induce modulated polar phases destabilizing the uniform N phase of such bend-shaped molecules (e.g. twist-bend or splay-bend), and can even induce polar blue phases [19,20].

The cybotactic clusters were first realized in the N phase of 1,2,4-oxadiazole based bent-core LCs [21]. It was detected through distinctive four-spot splitting observed in the SAXS (small angle X-ray scattering) pattern of the LC compounds. The cybotactic model of the N phase assumes the existence of short-range Sm-like ordering in the form of nanometer-sized clusters (cybotactic groups) exhibiting a layered supramolecular structure (typically skewed, i.e. SmC-like or sometimes orthogonal SmA-like) with intrinsic biaxial orientational order [2]. Such clusters are generated by frustration of the orientational/translational diffusion of bent-shaped molecules in the nematic phase developing a transverse correlation exaggerated by the bent-shape of the molecules [2]. Several other experimental techniques (e.g. dynamic light scattering (DLS), nuclear magnetic resonance (NMR) study, dielectric studies etc.) are also reported by researchers to exhibit similar results consistent with the $N_{Cyb}$ model, thus indicating the presence of Sm-like cybotactic clusters in the N phase of bent-core LCs [9,21,22]. Recently, cybotactic clusters in the N phase have been reported via Cryo-TEM investigations of the bent-core LC compounds [15]. The results show existence of these clusters typically in the length scales of 30-50 nm in the N phase of bent-core compounds, which is larger than those estimated via the positional correlation lengths.

Despite being verified experimentally, there are relatively few studies [23–26] on the theory of such clusters. Moreover, the theory and the experiments are not always in tandem [14,26,27]. While the experiments suggest that the clusters occupy only a small percentage (~ 3% of the LC molecules forming clusters) of the whole LC volume [14,27], no conclusive remarks have yet been established through the theoretical studies [23–26]. In a very recent work, Madhusudana has proposed a two-state model for the nematic phase of bent-core LCs and has studied the nematic-isotropic transition of a bent-core compound using the Landau-de Gennes (LdG) free energy in terms of two scalar order parameters $S_c$ and $S_g$ accounting for the clusters and the surrounding LC molecules respectively [26]. Based on a few simplified assumptions and through minimization of the LdG free-energy as a function of $S_c$ and $S_g$, they predict bulk values of these two scalar order parameters. The results show that a rather straightened (less-bent) excited state (ES) conformer of bent-core LCs form clusters with smectic type ordering. Although the model is able to predict clusters, it is limited in the sense that it assumes a fixed number of molecules inside a cluster which is temperature independent, does not include the effects of spatial inhomogeneities, director fields, boundary effects or the effects of confinement. Experiments, on the contrary, suggest a strong temperature dependency of the size and polarity of these clusters, and also accounts for the confinement and alignment conditions applied to the system [28]. Therefore, a sizable amount of theoretical work is still required for the foundation of the cybotactic nematic model in order to correctly predict the system dynamics and the experimental outcomes.

In this paper we have studied a 1D model of spatially inhomogeneous order parameters for the nematic phase of bent-core LCs in light of the earlier work by Madhusudana [26]. We introduce spatial inhomogeneities to the order parameter dependant free-energy and account for the gradient terms, which allows us to study the effects of boundary conditions and confinement. We use the same notations used by Madhusudana and denote the cluster order parameter as $S_c$ and the order parameter for the surrounding nematic LCs as $S_g$. We introduce, in addition, to the bulk free-energy, one-constant elastic energy density for $S_c$ and $S_g$ and have studied the associated Euler-Lagrange's equations subject to experimentally motivated boundary conditions. We have particularly focused on the effects of a parameter $C_2$ on the cluster order parameter ($S_c$) in our studies that accounts for the coupling between clusters and the surrounding LC molecules (γ) and the cluster size. We show that an enhanced coupling (γ) between the clusters and the surrounding LCs boosts the cluster order parameter ($S_c$). Experimentally, this suggests a strong core to core interaction (among the aromatic cores) between the constituent LCs and therefore an increase in the associated correlation length [28]. The enhanced coupling also suggests a transition to a weakly ordered phase when in the vicinity of nematic-isotropic transition. Moreover, we look at the electric-field analogous of the field-induced order parameters in line with our earlier experimental observations [28]. In section 2, we briefly describe the model, how it compares with the experimental conditions and the mathematical framework. In section 3, we discuss the numerical results and attempt to draw an analogy with the experimental findings followed by a conclusion of the work.

## 2. The Model

We consider the nematic phase of bent-core LC molecules in a confined geometry and assume that clusters are formed by straightened (less-bent) conformers of the LC molecules while still being surrounded by others that do not form clusters [26]. We denote the cluster molecules as ES (excited state) and the surrounding ones as GS (ground state), and their associated order parameters as $S_c$ and $S_g$ respectively [26]. Under usual experimental conditions, LC molecules are confined in a geometry of thickness typically in the range of few microns (μm) with a definite molecular alignment at the boundaries (substrate interfaces). The alignment condition is imposed by coating the two substrate surfaces with a polyimide layer (glass plates) and by gently rubbing the surfaces in a parallel or antiparallel fashion. This creates grooves on the polyimide layer that confines the LC molecular orientation at the surfaces and therefore further inside the sample LC volume. Therefore, formation of clusters at the boundaries is not a possibility and also the LCs near the boundary have a definite (fixed) orientation. Mathematically, this yields the Dirichlet boundary condition $S_c = 0$ at the two interfaces, which does not allow for clusters on the boundaries. Also, $S_g$ must have a constant non-zero value at the boundaries owing to the imposed orientation condition (aligning layer). We study the Landau-de Gennes (LdG) model for $S_c$ and $S_g$ proposed by Madhusudana and incorporate elastic energy densities that account for spatial inhomogeneities [26]. The corresponding Landau-de Gennes free energy is then written as,

$$F = K_1|\nabla S_g|^2 + K_2|\nabla S_c|^2 + (1-a_x)\left\{\frac{a_g}{2}(T-T^*)S_g^2 - \frac{B_g}{3}S_g^3 + \frac{C_g}{4}S_g^4 - E_{el}S_g\right\} + \frac{a_x}{N_c}\left\{-(1-a_x)\gamma S_g S_c + \frac{\alpha_c}{2}S_c^2 + \frac{\beta_c}{4}S_c^4\right\} - a_x J E_{el} S_c \quad (1)$$

Here, the '$g$' and '$c$' subscripts signify the Ground-state (GS) and the Excited-state (ES) molecules, respectively, and the cluster is essentially formed by the ES molecules. $K_1$ and $K_2$ denote the elastic constants, $S_g$ and $S_c$ represent the order parameters of the GS molecules and the clusters respectively. $a_g$, $B_g$, $C_g$, $T^*$ are the usual LdG parameters for the GS molecules with $\gamma$ being the constant of coupling between the GS molecules and the clusters [26]. $\alpha_c$, $\beta_c$ are coefficients for the saturation terms to ensure the absolute value of $S_c < 1$ and $N_c$ represents the number of ES molecules in each cluster. Moreover, the order parameters $S_g$ and $S_c$ are considered as the electric-field induced order parameters. $J$ accounts for the shape anisotropy of ES molecules, $E_{el}$ is the electric field energy ($=\frac{1}{2}\varepsilon_0\Delta\varepsilon E^2$) where $\varepsilon_0$ is the free-space permittivity, $\Delta\varepsilon$ is the dielectric anisotropy, $E$ is the applied electric field and $a_x$ is the mole fraction of the ES molecules given by

$$a_x = \exp(-E_{ex}/k_B T)/[1 + \exp(-E_{ex}/k_B T)] \quad (2)$$

where $E_{ex}$ denotes the excitation energy of the ES molecules, $k_B$ is the Boltzmann's constant and $T$ is the temperature.

The corresponding Euler-Lagrange's equation can be easily computed and the resulting coupled PDEs are,

$$2K_1\nabla^2 S_g = (1-a_x)[a_g(T-T^*)S_g - B_g S_g^2 + C_g S_g^3 - E_{el} - \frac{a_x \gamma S_c}{N_c}] \quad (3)$$

$$2K_2\nabla^2 S_c = a_x[\frac{-(1-a_x)\gamma S_g}{N_c} + \frac{\alpha_c S_c}{N_c} + \frac{\beta_c S_c^3}{N_c} - JE_{el}] \quad (4)$$

We now solve these coupled PDEs subject to the fixed boundary conditions $S_c = 0$ and $S_g =$ fixed constant (non-zero) at the two interfaces (*i.e.* at $x = 1$ and at $x = 0$ in 1D). We do not comment on the fixed value of $S_g$ at the interfaces except that it has to be non-zero to account for the imposed orientation at the boundaries.

## 2.1 1D Analogue

In 1D, the equations (3) and (4) can be written as,

$$2K_1 \frac{d^2 S_g}{dx^2} = (1 - a_x)[a_g(T - T^*)S_g - B_g S_g^2 + C_g S_g^3 - E_{el} - \frac{a_x \gamma S_c}{N_c}] \quad (5)$$

$$2K_2 \frac{d^2 S_c}{dx^2} = a_x[\frac{-(1 - a_x)\gamma S_g}{N_c} + \frac{\alpha_c S_c}{N_c} + \frac{\beta_c S_c^3}{N_c} - JE_{el}] \quad (6)$$

The fixed boundary conditions now read, (i) $S_g$ = constant (non-zero) and (ii) $S_c = 0$ at both $x = 0$ and at $x = 1$. This simply accounts for fixed nematic ordering and the absence of clusters at the boundaries.

### 2.1.1 Non-dimensionalization

To simplify the analysis a bit further, we non-dimensionalize the system (5) and (6) which allows for better comparison with experiments. Let us now consider $S_g = Y_1$, $S_c = Y_2$, $(1-a_x) a_g (T-T^*) = A$, $(1-a_x) B_g = B$, $(1-a_x) C_g = C$, $a_x (1-a_x) \gamma / N_c = D$, $(1-a_x) E_{el} = E$, $\alpha_c a_x / N_c = M$, $\beta_c a_x / N_c = N$ and $J E_{el} a_x = P$. From equations (5) and (6) we have,

$$2K_1 \frac{d^2 Y_1}{dx^2} = AY_1 - BY_1^2 + CY_1^3 - DY_2 - E \quad (7)$$

$$2K_2 \frac{d^2 Y_2}{dx^2} = MY_2 + NY_2^3 - DY_1 - P \quad (8)$$

It can be noticed that equations (7) and (8) are the Euler-Lagrange equations of the functional

$$\int_\Omega [\frac{A}{2}Y_1^2 - \frac{B}{3}Y_1^3 + \frac{C}{4}Y_1^4 - EY_1 - PY_2 + \frac{M}{2}Y_2^2 + \frac{N}{4}Y_2^4 - DY_1 Y_2 \\ + K_1 \left(\frac{dY_1}{dx}\right)^2 + K_2 \left(\frac{dY_2}{dx}\right)^2 ] \, dx \quad (9)$$

Where $\Omega = [0, 100]$.

Let $\tilde{x} = x/x_s$, $\tilde{Y}_1 = Y_1/Y_s$ and $\tilde{Y}_2 = Y_2/Y_s$ where $x_s$ and $Y_s$ are the scaling factors. Then from equations (7) and (8),

$$\frac{d^2\tilde{Y}_1}{d\tilde{x}^2} = \frac{Ax_s^2}{2K_1}\tilde{Y}_1 - \frac{Bx_s^2 Y_s}{2K_1}\tilde{Y}_1^2 + \frac{Cx_s^2 Y_s^2}{2K_1}\tilde{Y}_1^3 - \frac{Dx_s^2}{2K_1}\tilde{Y}_2 - \frac{Ex_s^2}{2K_1 Y_s} \quad (10)$$

$$\kappa \frac{d^2\tilde{Y}_2}{d\tilde{x}^2} = \frac{Mx_s^2}{2K_1}\tilde{Y}_2 + \frac{Nx_s^2 Y_s^2}{2K_1}\tilde{Y}_2^3 - \frac{Dx_s^2}{2K_1}\tilde{Y}_1 - \frac{Px_s^2}{2K_1 Y_s} \quad (11)$$

where, $\kappa = K_2/K_1 = 1$ (under one constant approximation $K_2 = K_1$). We set $\frac{|A|x_s^2}{2K_1} = 1$ and $\frac{Bx_s^2 Y_s}{2K_1} = 1$. This implies, $x_s = \sqrt{2K_1/|A|}$ and $Y_s = |A|/B$. Equations (10) and (11) can now be written as, for $A > 0$,

$$\frac{d^2\tilde{Y}_1}{d\tilde{x}^2} = \tilde{Y}_1 - \tilde{Y}_1^2 + C_1\tilde{Y}_1^3 - C_2\tilde{Y}_2 - C_3 \quad (12)$$

$$\frac{d^2\tilde{Y}_2}{d\tilde{x}^2} = C_4\tilde{Y}_2 + C_5\tilde{Y}_2^3 - C_2\tilde{Y}_1 - C_6 \quad (13)$$

for $A < 0$,

$$\frac{d^2\tilde{Y}_1}{\partial\tilde{x}^2} = -\tilde{Y}_1 - \tilde{Y}_1^2 + C_1\tilde{Y}_1^3 - C_2\tilde{Y}_2 - C_3 \quad (14)$$

$$\frac{d^2\tilde{Y}_2}{\partial\tilde{x}^2} = C_4\tilde{Y}_2 + C_5\tilde{Y}_2^3 - C_2\tilde{Y}_1 - C_6 \quad (15)$$

where $C_1 = \frac{C|A|}{B^2}$; $C_2 = \frac{D}{|A|}$; $C_3 = \frac{EB}{|A|^2}$; $C_4 = \frac{M}{|A|}$; $C_5 = \frac{N|A|}{B^2}$ and $C_6 = \frac{PB}{|A|^2}$. The equations are now non-dimensionalized and the new boundary conditions at $\tilde{x} = 100$ and at $\tilde{x} = 0$ are,

(i) $\tilde{Y}_1 = 1$
(ii) $\tilde{Y}_2 = 0$

The non-dimensionalized free energy is (tilde omitted),

$$\int_\Omega \left[ \frac{1}{2}\text{sgn}(A)Y_1^2 - \frac{1}{3}Y_1^3 + \frac{C_1}{4}Y_1^4 - C_3 Y_1 + \frac{C_4}{2}Y_2^2 + \frac{C_5}{4}Y_2^4 - C_6 Y_2 \right. \\ \left. - C_2 Y_1 Y_2 + \frac{1}{2}\left(\frac{dY_1}{dx}\right)^2 + \frac{\kappa}{2}\left(\frac{dY_2}{dx}\right)^2 \right] dx \quad (16)$$

We assume the parameters to be independent of temperature. Therefore, at a fixed temperature $T$ (and therefore at a fixed $a_x$), the parameter $C_1$ accounts for the LdG parameters of GS molecules and therefore essentially remains a constant, $C_2$ accounts for the ratio of coupling constant $\gamma$ and the number of ES molecules in a cluster ($N_c$), $C_3$ accounts for the electric field energy $E_{el}$, $C_4$ accounts for the ratio of cluster LdG parameter $\alpha_c$ and the product of $a_g$ and $N_c$ which means that it is primarily dependant on $N_c$ only, $C_5$ accounts for the ration $\beta_c a_g/B_g^2 N_c$ and therefore only depends on $N_c$, $C_6$ essentially accounts for the anisotropy factor $J$ and the electric field energy $E_{el}$. In our study, we will vary the parameter $C_2$ keeping all the other parameters ($C_1$ and $C_3$-$C_6$) fixed and study the effects on $S_c$ and $S_g$. Since we are considering the system at a constant temperature, this only describes the variation of the order parameters as a function of the coupling constant $\gamma$.

## 3. Numerical Results and Discussion

Using numerical simulations in C++ and COMSOL, we have calculated the values of the two order parameters $S_g$ and $S_c$ and have studied their variation with the parameter $C_2$. The two order parameters $S_g$ and $S_c$ are defined as the average of the second order Legendre polynomial accounting for the average molecular orientation in the GS molecules and in the clusters [29]. All other parameters ($C_1$ through $C_6$) apart from $C_2$ are kept constant. The variation of $C_2$ only accounts for the influence of the coupling constant $\gamma$ on the order parameters at a fixed temperature. We analyze two such cases, one for $A > 0$ and another for $A < 0$. Since $A = (1-a_x) a_g (T-T^*)$ and $a_x$ is smaller than 1, for $A > 0$ we must have $T > T^*$ and for $A < 0$ we will have $T < T^*$. The temperature $T > T^*$ is the high-temperature regime for which the isotropic phase is stable and the nematic (N) phase is stable for $T < T^*$. We use a non-dimensionalized 1D domain $\Omega = [0, 100]$ (corresponding to $\tilde{x}$) for the simulations. Since $x = \tilde{x} x_s$, and $x_s \sim 10^{-8} m$ (from the parameters used in our calculations), the non-dimensionalized 1D domain $0 \leq \tilde{x} \leq 100$ is in micron range, consistent with experiments. We have used the following parameter values in our calculations [26]: $K_2 = K_1 = K = 15$ pN $= 15*10^{-12}$ N $= 15*10^{-7}$ dyne (under one constant approximation); $a_g = 0.04$, $B_g = 1.7$, $C_g = 4.5$, $T^* = 355$ K, $T = 360$ K (for A > 0) and 350 K (for A < 0), $N_c = 50$, $\alpha_c = 0.22$, $\beta_c = 4.0$, $\gamma = 5.0$, $J = 1.2$, $E_{el} = 2000$ ergs/cm$^3$, $E_{ex} = 1.1*10^{-13}$ ergs ($a_g$, $B_g$, $C_g$, $\gamma$, $\alpha_c$ and $\beta_c$ in $10^7/4$ cgs units). This yields $A = 0.045*10^7$; $B = 0.3825*10^7$; $C = 1.0125*10^7$; $D = 0.00225*10^7$; $E = 1800$; $M = 0.0001*10^7$; $N = 0.002*10^7$ and $P = 240$ (in respective cgs units). Therefore we have, $C_1 = 0.31142$; $C_2 = 0.05$ (for $\gamma = 5$); $C_3 = 0.034$; $C_4 = 0.00222$; $C_5 = 0.000615$ and $C_6 = 0.00453$.

The parameter $C_2$ is varied in a range between 0.01 and 0.2 corresponding to γ in the range 1 to 20. The re-scaled order parameters $S_g$ and $S_c$ are not constrained to be less than unity. We first discuss the case $A > 0$, *i.e.* when $T > T^*$ and the system favours the isotropic phase ($T - T^* = 5$ K). The numerically computed solutions of $S_g$ and $S_c$ for $A > 0$ are shown in Figure 1. The cluster order parameter $S_c$ steadily increases with increasing $C_2$ in the bulk and reaches a maximum of $S_c \sim 10$ at $C_2 = 0.2$ (*i.e.* γ = 20). γ is a measure of the coupling strength between the clusters and the surrounding LC molecules. Therefore, an enhanced γ is expected to enhance the cluster order parameter ($S_c$) as it suggests a very strong core to core interaction (among the aromatic cores) between the constituent LCs and hence an increase in the associated correlation length [28]. The solutions to $S_g$ also exhibit an appreciable increase with increasing $C_2$ but with a distinct jump between $C_2 = 0.05$ and $C_2 = 0.1$ (at $C_2 \geq 0.07$ to be precise), indicating a transition to a more ordered phase, in the vicinity of nematic-isotropic transition. Therefore, an enhanced γ not only induces a strong tendency to form clusters but also invokes a stronger ordering in the surrounding molecules. This observation is very similar to that of the intermediate coupling induced paranematic phase reported by Madhusudana [26,30]. However, the values of $S_g$ are quite small compared to that of $S_c$, suggesting that clusters may exist in relatively disordered ambient phases and in the vicinity of the nematic-isotropic transition temperature. Our study also suggests that this intermediate phase can be further stabilized at higher temperatures (~5 K above $T^*$) when a stronger coupling (γ) is induced in the system. Experimentally, this may be realized by enhancing the transverse dipole moment of the LC molecules which will enhance the core to core interaction and therefore the formation of clusters [12].

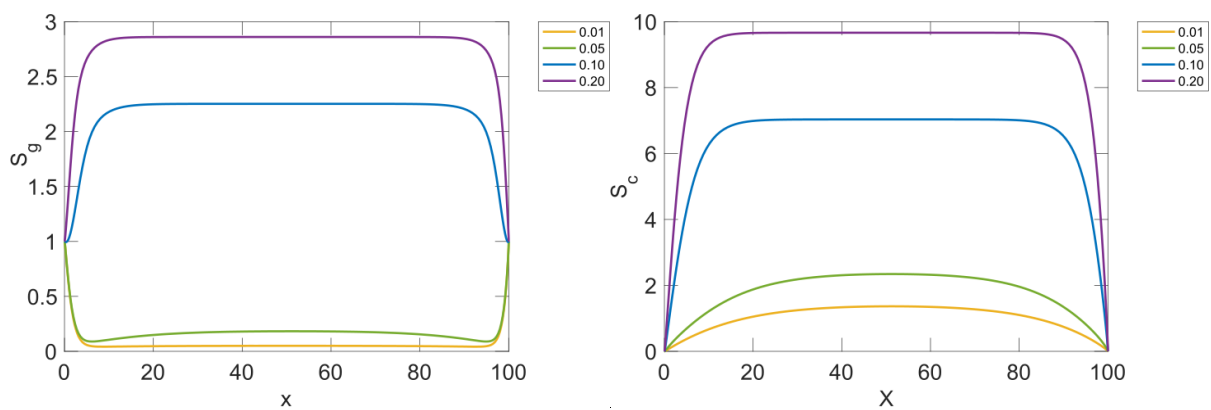

**Figure 1.** Numerical results for $S_g$ and $S_c$ in a domain $\Omega = [0, 100]$ with $C_2 = 0.01, 0.05, 0.1$ and $0.2$ (For $A = 0.045*10^7 > 0$).

When $A < 0$ (T = 350 K), the system is well within the nematic phase of LC. The numerical solutions for $S_g$ and $S_c$ for $A < 0$ are shown in Figure 2. As expected, the distinct jump in $S_g$ with increasing $C_2$ is not observed since the system is now in the nematic phase itself and does not undergo any transition. Only a small change in the $S_g$ value can be seen with increasing $C_2$ which shows that well inside the nematic phase, the nematic order is left unaltered by the enhanced coupling. In contrast, the cluster parameter $S_c$ is observed to change significantly with $C_2$ and thus with γ. This behaviour is similar to the case $A > 0$, but the values of $S_c$ are comparatively larger for $A < 0$ ($T < T^*$). This is again in line with the melting of clusters at higher temperatures [26]. Experimentally we have also observed similar signatures of reduction in cluster size with increasing temperature in one of our earlier works on bent-core systems [28]. Moreover, the $S_c$ profile, which is parabolic at smaller (~ 0.05) γ values (complying the presence of clusters near the bulk centre), becomes more flat and plateau-like at higher (~ 0.1 – 0.2) γ values suggesting that the clusters are evenly distributed in the bulk. It is interesting to note that the value of $S_g$ for $A > 0$ (when $T > T^*$) in case of an enhanced γ is comparable to $S_g$ for $A < 0$ for smaller values of γ (in the nematic phase), supporting our claim of a transition to a phase with nematic-like ordering (similar to the coupling induced paranematic phase) near the isotropic-nematic transition.

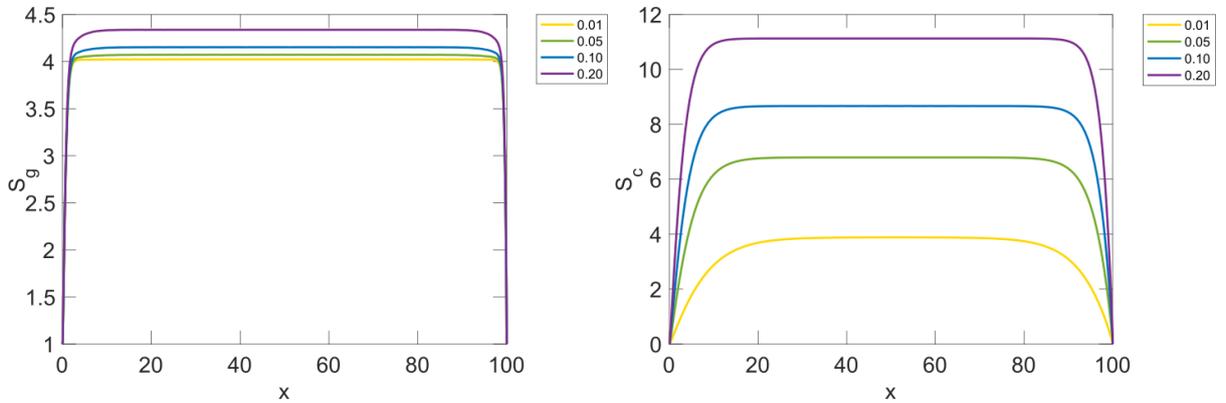

**Figure 2.** Numerical results for $S_g$ and $S_c$ in a domain $\Omega$ = [0, 100] with $C_2$ = 0.01, 0.05, 0.1 and 0.2 (For $A$ = -0.045*$10^7$ < 0).

We have also computed the maximum value of $S_c$ as a function of $C_2$ for both the cases $A < 0$ and $A > 0$. The variation of max($S_c$) with $C_2$ is shown in Figure 3. Both the profiles show a continuous increase in $S_c$ with $C_2$ due to the coupling induced ordering except for a sudden change when $C_2 \geq 0.07$ (*i.e.* γ ≥ 7) and $A > 0$. Again, the actual value of $S_c$ (= non-dimensionalized $S_c$ * scaling factor) is expected to be smaller than 1. Hence, there must be an upper bound on $C_2$ as well. The bound on max($S_c$) is captured by the dashed line in Figure 3.

For $A < 0$, the bound is $C_2 \leq 0.1$ while for $A > 0$ the bound is $C_2 \leq 0.16$. Therefore, we must have $\gamma \leq 10$ for physically relevant solutions in the nematic phase and $\gamma \leq 16$ in the high temperature regime.

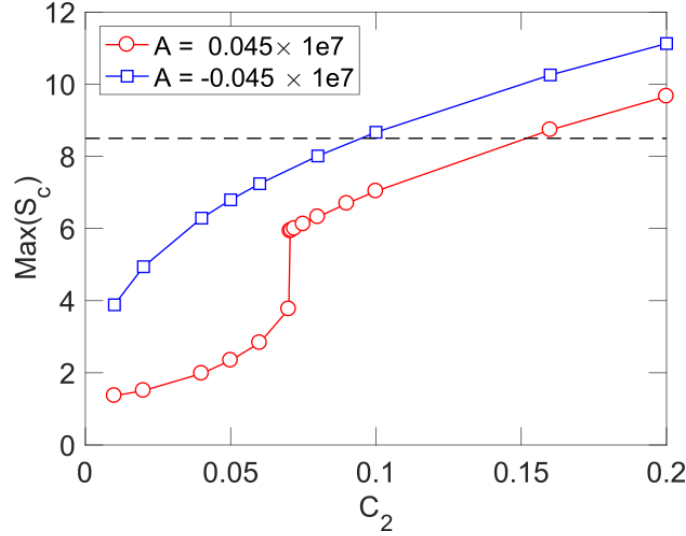

**Figure 3.** Upper bound of $S_c$ as a function of $C_2$ for both $A = 0.045*10^7$ ($> 0$) and $A = -0.045*10^7$ ($< 0$). The dashed line corresponds to $S_c = B/|A| = 8.5$, where $S_c$ is the re-scaled order parameter.

## 3.1 Preliminary work in 2D

In 2D, for $A > 0$ ($=0.045*10^7$) the Euler-Lagrange equations (12) and (13) are,

$$\frac{\partial^2 \widetilde{Y}_1}{\partial \widetilde{x}^2} + \frac{\partial^2 \widetilde{Y}_1}{\partial \widetilde{y}^2} = \widetilde{Y}_1 - \widetilde{Y}_1^2 + C_1 \widetilde{Y}_1^3 - C_2 \widetilde{Y}_2 - C_3 \tag{17}$$

$$\frac{\partial^2 \widetilde{Y}_2}{\partial \widetilde{x}^2} + \frac{\partial^2 \widetilde{Y}_2}{\partial \widetilde{y}^2} = C_4 \widetilde{Y}_2 + C_5 \widetilde{Y}_2^3 - C_2 \widetilde{Y}_1 - C_6 \tag{18}$$

We now solve equations (17) and (18) in a non-dimensionalized 2D domain $\Sigma = [0, 100 : 0, 100]$ with the boundary conditions (i) $S_g = 1$ and (ii) $S_c = 0$ at all four boundaries. The COMSOL generated plots for $S_g$ and $S_c$ when $C_2 = 0.01$ and $0.05$ are shown in Figure 4. It can be observed that the increment of $C_2$ has a similar effect on $S_c$ in 2D with very little effect on $S_g$ (as in 1D). Therefore, the coupling ($\gamma$) has a clear influence on the formation of clusters and effectively enhances the cluster order parameter $S_c$.

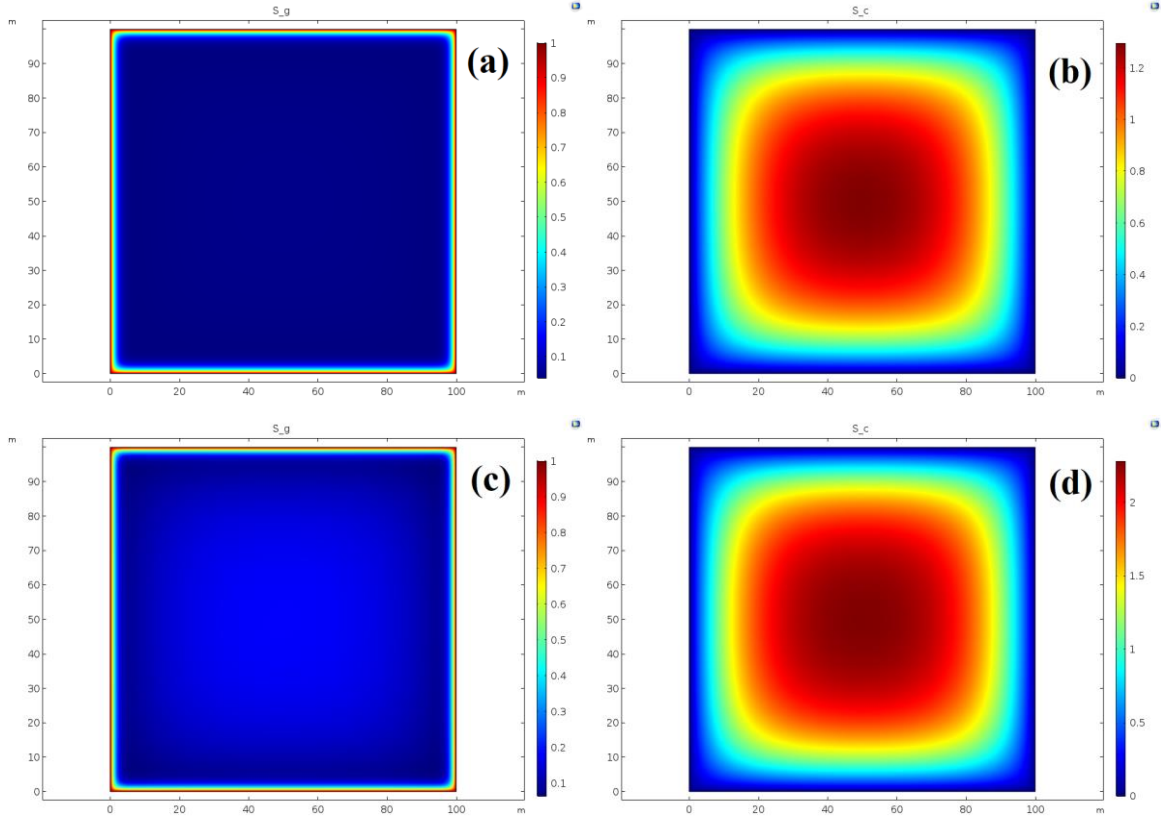

**Figure 4.** COMSOL generated solutions of (a), (c) $S_g$ and (b), (d) $S_c$ for $A > 0$ in a 2D domain $\Sigma = [0, 100 : 0, 100]$ when $C_2 =$ (a), (b) 0.01 and (c), (d) 0.05.

## 4. Conclusion

In summary, a 1D model of bent-core nematic system is studied in terms of the Landau-de Gennes free energy, the order parameters and the associated Euler-Lagrange's equations accounting for the spatial inhomogeneities and the effect of confinement. Although the system is non-dimensionalized, the scaling factors in our study matches well with that of the experiments and therefore the obtained results are experimentally significant. The study demonstrates the effect of coupling ($\gamma$) between the clusters and the surrounding LC molecules in the nematic as well as in the isotropic phase, very near to the nematic-isotropic transition. Increment in $\gamma$ is observed to enhance the cluster order parameter ($S_c$) and therefore signifying physical enhancement of cluster formation. The value of $S_c$ in the two cases ($T > T^*$ and $T < T^*$) also presents the effect of temperature on clusters with tendency to melt down/disappear at higher temperatures. Experimentally this can be related to the increased core to core interaction between the bent-core mesogens owing to their polar nature and promoting the formation of smectic-like highly ordered clusters in the nematic phase which extends into the isotropic phase [28]. Also, the $S_c$ profile, which was parabolic at smaller values of $\gamma$, becomes more flat and plateau-like when $\gamma$ is quite high (~ 0.1 – 0.2) suggesting that the clusters are evenly distributed in the bulk. We also observed a distinct jump in the maximum value of $S_c$ (Figure 3) for $A > 0$ and $C_2 \geq 0.07$. Accounting for the actual value of $S_c$ (= non-dim. $S_c$ * scaling factor), when $C_2 < 0.07$ (*i.e.* $\gamma < 7$) our results are comparable to that reported by Madhusudana (~ 0.6) [26], while for $C_2 \geq 0.07$ (*i.e.* $\gamma \geq 7$) and still in the range of acceptable solutions, we observe enhanced values of $S_c$ in the bulk (~ 0.9). Therefore, incorporating the spatial inhomogeneities as well as the effects of confinement and boundary conditions we are able to demonstrate an enhancement in the $S_c$ value which has significant experimental importance. $\gamma$ also has a profound effect on the order parameter $S_g$ for the surrounding molecules and shows a transition to a phase with weak nematic-like ordering when $T > T^*$. However, in the low temperature regime ($T < T^*$) the effect of $\gamma$ is only minute. The work is then further extended in 2D which supports the 1D observations. Even though the model is highly simplified, the obtained results have significant experimental importance and sheds light on the effect of coupling parameter $\gamma$ on the formation of clusters in the bent-core nematic systems. The study can also be improved by incorporating the aspects of director-field analysis which we plan to do as a part of our future work.